\pdfoutput=1
\documentclass[conference]{IEEEtran}
\usepackage{amsmath}
\usepackage{amsfonts}
\usepackage{booktabs}
\usepackage{graphicx}
\usepackage{hyperref}
\usepackage{cite}
\usepackage[utf8]{inputenc}
\usepackage[T1]{fontenc}

\makeatletter
\newcommand{\aicitenums}{[\@ifundefined{b@aiClaude}{?}{\csname b@aiClaude\endcsname}]--[\@ifundefined{b@aiBGE}{?}{\csname b@aiBGE\endcsname}]}
\makeatother
\newcommand{\aimark}{\textsuperscript{\dag}}
\newcommand{\aifirst}{\begingroup\renewcommand{\thefootnote}{\dag}%
  \footnote{AI-assisted content, in this and every section marked \dag;
  see the Acknowledgment~\aicitenums.}\addtocounter{footnote}{-1}\endgroup}

\newcommand{\pn}[2]{#2}

\IEEEoverridecommandlockouts

\makeatletter
\def\ps@IEEEtitlepagestyle{%
  \def\@oddhead{\parbox{\textwidth}{\centering\footnotesize
    This work has been submitted to the 2027 IEEE PES General Meeting for
    possible publication. Copyright may be transferred without notice, after
    which this version may no longer be accessible.}}%
  \let\@evenhead\@oddhead\def\@oddfoot{}\def\@evenfoot{}}
\makeatother
\begin{document}
\bstctlcite{IEEEexample:BSTcontrol}

\newcommand{\pending}[1]{\textbf{[pending: #1]}}

\title{Checking DER Commands Against Grid Context\\Catches Credentialed Attacks Static Checks Miss}

\author{\IEEEauthorblockN{Benjamin Blakely}
\IEEEauthorblockA{\textit{Argonne National Laboratory}\\
Lemont, IL, USA\\
bblakely@anl.gov}
\thanks{This work was supported in part by the U.S.\ Department of Energy,
Office of Cybersecurity, Energy Security, and Emergency Response (CESER),
through prime contractor Iowa State University. The submitted manuscript has
been created in part by UChicago Argonne, LLC, operator of Argonne National
Laboratory. Argonne, a DOE Office of Science laboratory, is operated under
Contract No.\ DE-AC02-06CH11357. The U.S.\ Government retains for itself, and
others acting on its behalf, a paid-up nonexclusive, irrevocable worldwide
license in said article to reproduce, prepare derivative works, distribute
copies to the public, and perform publicly and display publicly, by or on
behalf of the Government.}
}

\maketitle

\begin{abstract}
Authorization of distributed energy resource (DER) commands judges a command
by who sent it and whether its value has been seen before. An attacker holding
a legitimate credential who writes ordinary values passes both tests. We add a
check that judges each DER write against the grid's operating context: the bus
voltages around the target, the hour, the day-ahead price, system load, and
irradiance, compared with what operators wrote under similar conditions in the
feeder's own benign history. We evaluate it live on the IEEE 123-bus and
8500-node feeders, driven by a measured 2024 year of price, load, and
irradiance, against nine attack scenarios. A lookup trained on the same
history catches none of the writes of the four scenarios that send values
already familiar on the targeted device;
the context check refuses \pn{DetectInsiderCredentialAbuseTsevenPct}{91.7\%} of
insider-abuse writes and \pn{DetectDerAggregationTsevenPct}{64.4\%} of
DER-aggregation writes while denying \pn{BenignTsevenPct}{1.28\%} of benign
writes. In an offline test that replays recorded moments with curtailment
setpoints drawn from the benign range, issued at a positive price, it refuses
\pn{CurtailIeeeOneTwentyThreeContextPct}{47.3\%} and
\pn{CurtailIeeeEightyFiveHundredContextPct}{50.5\%} on the two feeders, against
\pn{CurtailIeeeOneTwentyThreeValueOnlyPct}{18.5\%} and
\pn{CurtailIeeeEightyFiveHundredValueOnlyPct}{16.0\%} for the same check
without context; live, the few setpoints tested were mostly shallow and were
mostly allowed. The context check complements the lookup.
\end{abstract}

\begin{IEEEkeywords}
distribution system security, distributed energy resources, authorization,
zero trust, situational awareness
\end{IEEEkeywords}

\section[Introduction]{Introduction\aifirst}
\label{sec:intro}

Coordinated manipulation of many DERs can cause instability, unnecessary protection tripping, or blackouts beyond the impact of any single compromised asset~\cite{qiCybersecurityDistributedEnergy2016,
chenCybersecurityDistributedEnergy2025,
sugunarajDistributedEnergyResource2026}. The commands involved need not look
wrong. Static authorization judges a command by who sent it, what it targets,
and whether its value resembles values seen before. An adversary holding a
distributed energy resource management system (DERMS) credential, or an
insider, can write each photovoltaic (PV) unit in a zone using the expected source and register to send a value operators routinely write, and passes every one of these tests.

What such writes get wrong is the moment. Where operators curtail PV when
midday solar drives the price negative, a deep curtailment setpoint is an
ordinary value on that channel, yet a control-room operator would refuse the
same setpoint at noon on a day the price is positive and voltages are normal. The judgment concerns
consistency between a command and the conditions it is issued in, and such a
curtailment need not violate any physical limit, so safety-based command
validation need not refuse it.

Our authorization check makes this judgment automatically. It learns from a feeder's benign history how operators set each device under
given conditions and refuses writes whose value does not fit. Deployed in the authorization path of a Modbus testbed and compared with a lookup learned from the same history, it catches attacks the lookup misses at a fraction of the lookup's false denials, and a pre-registered test with ablations measures whether context catches routine curtailment values issued at a positive price.

\section[Background]{Background\aimark}
\label{sec:background}

Coordinated manipulation of many individually valid loads is an established attack class~\cite{aminiDynamicLoadAltering2018,
dabrowskiGridShockCoordinated2017, malekiSurveyLoadAlteringAttacks2025}, and coordinated DER operation through aggregators is expected to amplify the impact of single-asset compromise~\cite{chenCybersecurityDistributedEnergy2025}.

Zero-trust architectures for grid and operational technology networks
assess trust continuously~\cite{jamesZeroTrustArchitecture2023,
chenTrustAssessmentDistributed2026, kapoorCyberSecurityOT2025}, but deployed grid security is organized around perimeters and zones, as in NERC
CIP-005~\cite{nercCIP0057}
and IEC
62443-3-3~\cite{iec62443-3-3}.
A zone authorizes requester--target pairs, not commands under current conditions.

Context-aware schemes update a requester's trust from measurement-based intrusion alerts or the time of request~\cite{fragkosDynamicRoleBasedAccess2022,
jamesZeroTrustArchitecture2023}, or gate access to supervisory control (SCADA) data on conditions
such as active alarms~\cite{routrayContextAwareAttributeBased2023}, and DER
security analyses call for access decisions that change with grid and DER
state~\cite{qiCybersecurityDistributedEnergy2016}. In these, context changes requester trust or data access; it does not judge each command before it executes.
Command validation executes
each incoming command in a model of the current grid state and forwards it only
if the predicted state is consistent and
safe~\cite{mashimaCybersecurityModernSmart,
chromikIntegratedTestbedLocally2018}, or checks received commands against local
measurements~\cite{universityofillinoisCyberResilientFlexible2022}. These test
physical consistency and safety, which a curtailment issued at the wrong price
does not violate. Exogenous data has been used in detection, chiefly
irradiance and weather to expose falsified DER generation
readings~\cite{krishnaEvaluatingDetectorsOptimal2018,
ismailDeepLearningDetection2020}.

Closest in spirit, inverter measurements have been used to detect
manipulated volt-VAR curves~\cite{saberLearningBasedDetectionMalicious2023};
that detector judges a received curve from the inverter's response once the curve is applied, whereas the context check decides in the authorization path before a write reaches
the device.

Frequent false alarms lead operators to ignore and distrust a detector~\cite{brandbyManagementAlarmSystems1997,
koistinenSpatioTemporalAttentionGraph2026}, and a false denial in an
authorization path withholds a control action outright, so detection gains must be weighed against false denials.

\section[Method]{Method\aimark}
\label{sec:method}

\subsection{Threat model}
\label{sec:threat}

The adversary controls the identity of one or more remote terminal units (RTUs) or the DERMS
credential, by theft, an insider, or a firmware compromise, and issues Modbus
writes under that identity through the same MTU path a legitimate command
takes. Every write uses a legal function code, targets a register the identity
may write, and carries a value inside the register's range. The adversary
knows the feeder's asset inventory and can read the public day-ahead price. It
cannot alter telemetry, the price feed, the policy engine, or the checks'
fitted models; false data injection on measurements is outside this study.

The primary scenario is \emph{price-timed curtailment}. The
DERMS credential curtails every PV unit in one zone to a single fraction of
nameplate. The fraction is drawn once per cell, with a fixed seed, from that
cell's own benign DERMS PV setpoints below 0.95 of nameplate in the fitting
data, so the commanded value is one operators have written on that feeder. The
attack waits for the first hour between 10:00 and 16:00 local standard time,
with at least six simulated hours before dark, whose day-ahead price is
non-negative, writes the whole zone at once, and re-asserts the setpoints four
times. It removes generation the market is paying for, with a value benign
operators write when the price is negative.

Three send values operators also send on the targeted device. \emph{DER aggregation} sets every DER in
one RTU's zone to 0\,kW in three daylight windows and holds it;
\emph{insider abuse} sends the same setpoints as an off-cadence burst;
\emph{firmware cohort} curtails the DERs sharing a vulnerable firmware version
across two or three zones, each through its own zone's RTU. \emph{Regulator
aggregation} holds a zone's regulators at 0.95\,pu. Three open
breakers (credential outage, ransomware spillover, and a long-staged intrusion
with one cross-zone open), and one probes safety controllers with a
read/write-multiple-registers request (function code 23), which benign
traffic never sends, before writing to them.

\subsection{Testbed}
\label{sec:testbed}

We use the IEEE 123-bus and IEEE 8500-node OpenDSS feeder models distributed
with PowerGym~\cite{fanPowerGymReinforcementLearning2022}, with DER overlays: a
small feeder whose every bus is observed and a large one that is only partly
observed.
The 123-bus feeder has \pn{BusesIeeeOneTwentyThree}{157} buses,
\pn{DerPvIeeeOneTwentyThree}{25} photovoltaic DERs,
\pn{DerBessIeeeOneTwentyThree}{3} batteries and
\pn{RegulatorsIeeeOneTwentyThree}{7} regulators; the 8500-node feeder has
\pn{BusesIeeeEightyFiveHundred}{4896} buses, \pn{DerPvIeeeEightyFiveHundred}{20}
photovoltaic DERs and \pn{RegulatorsIeeeEightyFiveHundred}{12} regulators.
Telemetry reads \pn{GridStateObservedBusesIeeeOneTwentyThree}{157} and
\pn{GridStateObservedBusesIeeeEightyFiveHundred}{1000} of those buses; the
voltage register block holds 1000 buses, filled in the model's bus order.
Each feeder is cut into \pn{ZonesIeeeOneTwentyThree}{5} zones: a breadth-first
tree is rooted at the substation, and subtrees whose sizes are closest to an
equal share are cut off in turn, so every zone is connected. One RTU serves each
zone, and a device belongs to the zone of its host bus.

A power-flow simulator
solves the feeder in OpenDSS and exposes DER limits, regulator setpoints,
breakers and bus voltages as Modbus registers. A traffic generator issues
every benign write to a master terminal unit (MTU), which asks an Open Policy
Agent (OPA) policy engine whether to forward it. The policy denies by default.
For each write it queries the grid-context check, a separate service, and
denies the write if that service fails or does not answer in time. Attack
writes enter through the same MTU, and every decision is logged.

Power flow runs as a quasi-static time series in OpenDSS daily mode, with one
solve per wall-clock second and simulated time running 2880 times faster than
wall time on the 123-bus feeder and 960 times faster on the 8500-node feeder,
whose slower solves would otherwise stretch an attack past dark. Benign roles
issue commands per wall-clock second, so the 8500-node feeder carries about
three times as many benign writes per simulated hour. PV ratings and loads
are scaled per seed by factors drawn from $[0.97, 1.03]$ and $[0.95, 1.05]$.

The simulator holds a 2024 data year for the Fresno, California
area at hourly resolution, indexed to local standard time so that hour 12 is
near solar noon: global horizontal irradiance from the Open-Meteo historical
archive~\cite{zippenfenigOpenMeteocomWeatherAPI2024}, and the day-ahead
locational marginal price (LMP) at the California ISO (CAISO) ZP26 trading hub
and actual CAISO system load from CAISO OASIS~\cite{caisoOASIS2024}. Irradiance scales PV output. Every
feeder load follows one of the three year-long load shapes distributed with the
123-bus model, so feeder load comes from a different source year than price
and irradiance. The substation voltage is offset from its standard setpoint
by up to 0.02\,pu, falling as normalized system load $z$ rises. Simulated time starts on 1 April 2024, so the
baseline falls in April and May, when midday prices are often negative and PV
output is high against load. Attack episodes follow from May to early July on
the 123-bus feeder and from May to mid-August on the 8500-node feeder; the
price-timed attack fires in June on every cell, at the first positive 10:00
price after the preceding episode.

\subsection{Benign operation}
\label{sec:benign}

Seven simulated roles generate benign traffic: a DERMS writing about eight
setpoints per simulated hour, a volt-VAR optimization (VVO) controller, and five
human roles, among them a dispatcher with three sessions a day; human sessions
start at random times at a fixed average rate. Each DERMS setpoint is the smaller of two
limits. The first is an IEEE 1547-2018 volt-watt characteristic~\cite{IEEE1547ApplicationGuide2024} applied to the DER's last
measured bus voltage: full output at or below
1.05\,pu, falling linearly to zero at 1.09\,pu, read through 0.002\,pu Gaussian
measurement noise. The second is a price limit: full nameplate at a
non-negative LMP, and a fraction $\max(0, 1 + \lambda/50)$ of nameplate at a
negative LMP $\lambda$ in \$/MWh, so a unit is fully curtailed at
$-50$\,\$/MWh. Curtailed setpoints are
therefore routine in negative-price hours. The VVO controller steps a
regulator when its voltage leaves $[0.955, 1.045]$\,pu, dispatcher load
setpoints fall as system load rises, and other human roles open and close
breakers for planned switching. Benign writes from outside the target's zone are excluded from the false-denial rate.

Before any attack is run, each feeder's baseline simulation must show that operators
respond to these conditions, with at least a share 0.10 of DERMS setpoints
below 0.95 of nameplate and a setpoint--LMP Spearman correlation of at least
0.3. Both feeders pass:
\pn{OperatorIeeeOneTwentyThreeDerBelowNinetyFivePct}{44.6\%} and
\pn{OperatorIeeeEightyFiveHundredDerBelowNinetyFivePct}{38.9\%} of setpoints
are curtailed, with correlations of \pn{OperatorIeeeOneTwentyThreeDerLmpSpearmanRho}{0.75}
and \pn{OperatorIeeeEightyFiveHundredDerLmpSpearmanRho}{0.82}.

\subsection{Authorization checks}
\label{sec:checks}

A protocol check admits only legal function codes from the trusted network, and
a role-and-zone check (the \emph{zone check}) requires a recognized
requester role and a target inside the requester's zone. The grid-context check
is added on top of these and applied to the write function codes.

\emph{Grid context.} For a write $w$ of value $v_w$ to asset $a$, the check
forms a feature vector $x_w$ of ten values. Five are bus voltages read from the
simulator: $a$'s own bus, the mean, maximum and minimum over observed buses
within two hops of $a$, and the feeder maximum.
The other five describe the operating moment:
the hour of day $h$ as $\sin(2\pi h/24)$ and $\cos(2\pi h/24)$, the LMP, the
CAISO system load, and the irradiance, all taken from the data year's row for
the write's timestamp; none is computed from the write. Among the baseline writes to
assets of the same type, let $\mathcal{N}_k(x_w)$ be the $k$ whose features are
nearest to $x_w$ in Euclidean distance after per-type standardization. The
score is
\begin{equation}
s(w) = \Bigl|\, \frac{v_w}{c_a} - \operatorname*{median}_{j \in \mathcal{N}_k(x_w)} \frac{v_j}{c_j} \Bigr|,
\label{eq:score}
\end{equation}
with $k=\pn{GridStateNeighbours}{25}$ and $c_a$ the largest value written to
$a$'s channel in the fitting data. The write is refused when $s(w)$ exceeds its
asset type's threshold $\tau$. The check therefore asks how far the value lies
from what operators wrote to similar devices under similar conditions; it
encodes no operator rule.

Thresholds are set on held-out baseline. The neighbours are drawn from all but
the last \pn{GridStateHoldoutDays}{7} baseline days, and each type's $\tau$
starts at the $1-\beta$ quantile of its held-out scores, with
$\beta=\pn{GridStateBudgetPct}{1\%}$, raised until the held-out refusal
rate over all types is within $\beta$; no threshold is set below 1.5 times a
channel's smallest value step, so one regulator tap never triggers a refusal. A
write to an asset type with fewer than 200 baseline writes, or to an asset with
no observed bus within two hops, is \emph{unmodelled} and passes; a write whose
own-bus voltage cannot be read is refused.

\emph{Ablations.} Two variants of the same check separate what each group of
features contributes. The \emph{value-only} variant gives every fitting write
of a type the same feature vector, so its score is the distance from the
median of all of that type's baseline values; it refuses values that are rare
on the channel regardless of conditions. The \emph{voltage} variant uses the
five voltages and drops the five context features. Both are fitted on the same
writes with the same holdout, floors, eligibility and budget as the live check,
and are replayed offline over the live configuration's traffic.

\emph{Lookup.} A static lookup learned from the same baseline checks familiarity. Its first test asks whether
the requesting device commonly writes to the target asset: it counts writes
for each device--asset pair and smooths the counts with a rank-3 truncated
singular value decomposition (SVD), so a pair never seen in the baseline
still scores well if similar pairs are common. The second refuses a value outside the channel's
baseline range, widened by a
margin. Both share the budget of \pn{LookupBudgetPct}{1\%} on the held-out
week. The lookup is replayed offline, alone and with the grid-context check; the combination refuses a write if either check does.

\subsection{Protocol and metrics}
\label{sec:protocol}

A \emph{cell} is one feeder and seed. Each cell runs
\pn{BaselineSimDays}{30} simulated days of benign traffic, the baseline on which
every learned check is fitted, followed by one attack episode per scenario,
in an order shuffled per seed. Benign traffic continues through every episode,
and the stack is not restarted between episodes, so an attack meets the state
left by the episodes before it. Every attack write is labelled adversarial at
the source. We run five seeds per feeder with the grid-context check live.

A cell is excluded if its bus telemetry freezes or a normally energized bus
stays dark in the baseline, or if it lacks exactly one labelled episode per
scenario, a rule fixed before any result was read. Cells that failed through
harness faults (memory, a request timeout, and a simulation rate too fast for
the larger feeder) were rerun after each fault was fixed.
All \pn{CellsIncludedGridStateIeeeOneTwentyThree}{5} cells of the 123-bus and
\pn{CellsIncludedGridStateIeeeEightyFiveHundred}{5} of the 8500-node feeder
pass; they contain \pn{WritesAttackerGridState}{11257} attack writes and
\pn{WritesBenignInZoneGridState}{34951} benign in-zone episode writes.

Let $A$ be the attacker writes and $B$ the benign writes the zone check allows.
Detection is $|\{w \in A : w \text{ refused}\}|/|A|$, and the false-denial rate
is $|\{w \in B : w \text{ refused}\}|/|B|$. A false denial withholds a legitimate
control action, so it measures an availability cost. Rates pool writes over
cells, and intervals are \pn{BootstrapLevelPct}{95\%} percentile bootstrap
intervals over \pn{BootstrapResamples}{10000} resamples of whole cells, since
episodes within a cell share a running stack and are not independent.

Whether the check refuses a curtailment depends on how deep it is, and a live
cell's attack repeats a single randomly drawn value, so five cells per feeder
sample only five depths. To measure the check across the full range of
curtailment values operators use, the price-timed attack is also tested
offline. Context features other than voltage depend only on the timestamp, so
recorded moments can be replayed. Every benign PV write outside the fitting
data, in daylight (irradiance at least 200\,W/m$^2$) and to a modelled target,
becomes a probe: its value is replaced by one drawn at random from the cell's
benign PV setpoints below 0.95 of the channel maximum, and its recorded
features are kept. Probes at a non-negative price are the attack; probes at a
negative price, when operators curtail too, are the control and approximate
an attacker who times to the public price. Each probe is scored by the
grid-context check and both ablations. Harm is the curtailment a probe would
cause, in units of nameplate: below nameplate at a non-negative price, and
below the price fraction $\max(0, 1 + \lambda/50)$ at a negative one. A
replayed probe keeps the voltages recorded before it; a real curtailment would
lower them.

A criterion fixed before any probe was scored requires that on each feeder,
with at least 200 non-negative-price probes, the grid-context check must
refuse at least 20 percentage points more of them than the value-only variant,
while its false-denial rate on benign episode writes stays at or below 0.02.

\section[Results]{Results\aimark}
\label{sec:results}

\subsection{Detection and false denial}

\begin{table}[t]
\centering
\caption{Attack writes refused, pooled over both feeders and ten cells, by
scenario and by target. Every configuration includes the zone check.}
\label{tab:detection}
\footnotesize
\setlength{\tabcolsep}{4pt}
\begin{tabular}{@{}lrrr@{}}
\toprule
Scenario & Lookup & Context & Both \\
\midrule
Insider abuse & \pn{DetectInsiderCredentialAbuseLookupPct}{0.0\%} & \pn{DetectInsiderCredentialAbuseTsevenPct}{91.7\%} & \pn{DetectInsiderCredentialAbuseBothPct}{91.7\%} \\
DER aggregation & \pn{DetectDerAggregationLookupPct}{0.0\%} & \pn{DetectDerAggregationTsevenPct}{64.4\%} & \pn{DetectDerAggregationBothPct}{64.4\%} \\
Firmware cohort & \pn{DetectFirmwareCohortExploitationLookupPct}{0.0\%} & \pn{DetectFirmwareCohortExploitationTsevenPct}{56.2\%} & \pn{DetectFirmwareCohortExploitationBothPct}{56.2\%} \\
Price-timed curtail. & \pn{DetectPositivePriceCurtailmentLookupPct}{0.0\%} & \pn{DetectPositivePriceCurtailmentTsevenPct}{3.9\%} & \pn{DetectPositivePriceCurtailmentBothPct}{3.9\%} \\
Regulator aggregation & \pn{DetectRegulatorAggregationLookupPct}{100.0\%} & \pn{DetectRegulatorAggregationTsevenPct}{84.2\%} & \pn{DetectRegulatorAggregationBothPct}{100.0\%} \\
Credential outage & \pn{DetectCoordinatedCredentialOutageLookupPct}{88.2\%} & \pn{DetectCoordinatedCredentialOutageTsevenPct}{0.0\%} & \pn{DetectCoordinatedCredentialOutageBothPct}{88.2\%} \\
Ransomware spillover & \pn{DetectRansomwareSpilloverLookupPct}{100.0\%} & \pn{DetectRansomwareSpilloverTsevenPct}{0.4\%} & \pn{DetectRansomwareSpilloverBothPct}{100.0\%} \\
Long-staged destr. & \pn{DetectLongStagedDestructiveLookupPct}{95.0\%} & \pn{DetectLongStagedDestructiveTsevenPct}{25.0\%} & \pn{DetectLongStagedDestructiveBothPct}{95.0\%} \\
Safety bypass & \pn{DetectSafetySystemBypassLookupPct}{100.0\%} & \pn{DetectSafetySystemBypassTsevenPct}{100.0\%} & \pn{DetectSafetySystemBypassBothPct}{100.0\%} \\
\midrule
DER/regulator writes & \pn{DetectValueTargetsLookupPct}{12.0\%} & \pn{DetectValueTargetsTsevenPct}{67.8\%} & \pn{DetectValueTargetsBothPct}{69.5\%} \\
Breaker writes & \pn{DetectBreakerTargetsLookupPct}{94.2\%} & \pn{DetectBreakerTargetsTsevenPct}{0.1\%} & \pn{DetectBreakerTargetsBothPct}{94.2\%} \\
Benign writes & \pn{BenignLookupPct}{16.26\%} & \pn{BenignTsevenPct}{1.28\%} & \pn{BenignBothPct}{16.83\%} \\
\bottomrule
\end{tabular}
\end{table}

The two learned checks catch different attacks (Table~\ref{tab:detection}).
The lookup refuses most attack writes, including every regulator and
breaker scenario, because those attacks write pairs or values the baseline
never held. It refuses none of the writes of the four scenarios that send
familiar values from the expected device. The grid-context check refuses most
insider-abuse writes and more than half of the DER-aggregation and
firmware-cohort writes. It does not model breakers, so it passes the breaker
writes that make up most attack writes, and its rate over all attack writes
(\pn{DetectAllTsevenPct}{11.1\%}) mainly reflects that scenario mix. Combined,
the two refuse
\pn{DetectAllBothPct}{90.4\%} of attack writes
(\pn{DetectAllBothLoPct}{80.3\%}--\pn{DetectAllBothHiPct}{96.7\%}).

The two differ more in cost. The grid-context check denies
\pn{BenignTsevenPct}{1.28\%} (\pn{BenignTsevenLoPct}{0.81\%}--\pn{BenignTsevenHiPct}{1.71\%})
of benign in-zone writes, close to its budget, and the lookup
\pn{BenignLookupPct}{16.26\%} (\pn{BenignLookupLoPct}{13.06\%}--\pn{BenignLookupHiPct}{20.16\%}),
\pn{BenignIeeeEightyFiveHundredLookupPct}{23.16\%} on the 8500-node feeder.
Both were calibrated to the same budget on the held-out week; only the
context check stays near it in the weeks that follow. A one-percent budget can still deny many commands on a busy feeder and is unlikely to be deployable unchanged; it serves here as a common operating point for comparing the checks, and deployment would need its own calibration.

Coverage limits the context check on the partly observed feeder. On the
8500-node feeder
\pn{CoverageValueAttackUnmodelledIeeeEightyFiveHundredPct}{26.6\%} of attack
writes to DERs and regulators target a device with no observed bus within two
hops; the check passes these unjudged and refuses
\pn{DetectUnmodelledIeeeEightyFiveHundredTsevenPct}{0.0\%} of them, against
\pn{DetectModelledIeeeEightyFiveHundredTsevenPct}{84.5\%} of writes to modelled
devices. Every device on the 123-bus feeder is modelled.

\subsection{Curtailment at a positive price}

Live, the grid-context check refused
\pn{DetectIeeeOneTwentyThreePositivePriceCurtailmentTsevenPct}{6.5\%} of
price-timed curtailment writes on the 123-bus and
\pn{DetectIeeeEightyFiveHundredPositivePriceCurtailmentTsevenPct}{0.0\%} on the
8500-node feeder. Each cell repeats one drawn value, and the draws were mostly
shallow curtailments; on the 123-bus feeder only the deepest of the five was
refused.

\begin{table}[t]
\centering
\caption{Offline curtailment probes: share refused, and curtailment passed per
attempted write in units of nameplate. Attack: probes at a non-negative price,
when operators do not curtail. Control: probes at a negative price, when
operators curtail too.}
\label{tab:curtail}
\footnotesize
\setlength{\tabcolsep}{4pt}
\begin{tabular}{@{}lrrrr@{}}
\toprule
& \multicolumn{2}{c}{Attack} & \multicolumn{2}{c}{Control} \\
\cmidrule(lr){2-3}\cmidrule(l){4-5}
& Refused & Passed & Refused & Passed \\
\midrule
\multicolumn{5}{@{}l}{\emph{123-bus}} \\
Value only & \pn{CurtailIeeeOneTwentyThreeValueOnlyPct}{18.5\%} & \pn{CurtailIeeeOneTwentyThreeHarmPassedValueOnly}{0.36} & \pn{CurtailIeeeOneTwentyThreeControlValueOnlyPct}{18.4\%} & \pn{CurtailIeeeOneTwentyThreeControlHarmPassedValueOnly}{0.13} \\
Voltage & \pn{CurtailIeeeOneTwentyThreeVoltagePct}{15.5\%} & \pn{CurtailIeeeOneTwentyThreeHarmPassedVoltage}{0.39} & \pn{CurtailIeeeOneTwentyThreeControlVoltagePct}{13.5\%} & \pn{CurtailIeeeOneTwentyThreeControlHarmPassedVoltage}{0.16} \\
Context & \pn{CurtailIeeeOneTwentyThreeContextPct}{47.3\%} & \pn{CurtailIeeeOneTwentyThreeHarmPassedContext}{0.17} & \pn{CurtailIeeeOneTwentyThreeControlContextPct}{25.2\%} & \pn{CurtailIeeeOneTwentyThreeControlHarmPassedContext}{0.12} \\
\multicolumn{5}{@{}l}{\emph{8500-node}} \\
Value only & \pn{CurtailIeeeEightyFiveHundredValueOnlyPct}{16.0\%} & \pn{CurtailIeeeEightyFiveHundredHarmPassedValueOnly}{0.37} & \pn{CurtailIeeeEightyFiveHundredControlValueOnlyPct}{15.0\%} & \pn{CurtailIeeeEightyFiveHundredControlHarmPassedValueOnly}{0.15} \\
Voltage & \pn{CurtailIeeeEightyFiveHundredVoltagePct}{37.8\%} & \pn{CurtailIeeeEightyFiveHundredHarmPassedVoltage}{0.22} & \pn{CurtailIeeeEightyFiveHundredControlVoltagePct}{31.8\%} & \pn{CurtailIeeeEightyFiveHundredControlHarmPassedVoltage}{0.09} \\
Context & \pn{CurtailIeeeEightyFiveHundredContextPct}{50.5\%} & \pn{CurtailIeeeEightyFiveHundredHarmPassedContext}{0.15} & \pn{CurtailIeeeEightyFiveHundredControlContextPct}{29.5\%} & \pn{CurtailIeeeEightyFiveHundredControlHarmPassedContext}{0.10} \\
\bottomrule
\end{tabular}
\end{table}

On the 123-bus feeder the grid-context check
refuses
\pn{CurtailIeeeOneTwentyThreeContextPct}{47.3\%}
(\pn{CurtailIeeeOneTwentyThreeContextLoPct}{42.2\%}--\pn{CurtailIeeeOneTwentyThreeContextHiPct}{52.0\%})
of \pn{CurtailIeeeOneTwentyThreeArmAProbes}{5037} attack probes,
\pn{CurtailIeeeOneTwentyThreeGapPts}{28.8} points more than the value-only
variant, at a false-denial rate of \pn{CurtailIeeeOneTwentyThreeFalseDenialPct}{1.49\%}.
On the 8500-node feeder it refuses \pn{CurtailIeeeEightyFiveHundredContextPct}{50.5\%}
(\pn{CurtailIeeeEightyFiveHundredContextLoPct}{42.5\%}--\pn{CurtailIeeeEightyFiveHundredContextHiPct}{60.4\%})
of \pn{CurtailIeeeEightyFiveHundredArmAProbes}{2780}, a gap of
\pn{CurtailIeeeEightyFiveHundredGapPts}{34.5} points, at
\pn{CurtailIeeeEightyFiveHundredFalseDenialPct}{0.47\%}. The pre-registered
criterion is met on both feeders, and the context check passes less than half
the curtailment the value-only variant passes (Table~\ref{tab:curtail}). The value-only variant refuses attack and control probes at nearly the
same rate, since both draw from the same values; the context check refuses
attack probes more often than control probes. At a negative price, the
context check passes less curtailment than the value-only variant on both
feeders, though on the 8500-node feeder the voltage-only ablation passes less
still.

\section[Discussion]{Discussion\aimark}
\label{sec:discussion}

For the context check to work in practice, two conditions
follow from the measurements. First, the targets must be observed: on the
8500-node feeder a quarter of value-attack writes went to devices the check
could not judge, and it refused none of them. Second, the curtailment must be
deep enough to stand out; live, the shallow curtailments drawn were mostly
allowed, so the check reduces price-timed curtailment without stopping it.

How much the non-voltage features add depends on the feeder. On the 8500-node feeder the voltage-only ablation refuses
\pn{CurtailIeeeEightyFiveHundredVoltagePct}{37.8\%} of attack probes against the full check's
\pn{CurtailIeeeEightyFiveHundredContextPct}{50.5\%}; on the 123-bus feeder it refuses fewer than the
value-only variant, so the gain there comes from hour, price, load and
irradiance together.

The feeders, Modbus traffic and rule-based operators are simulated from one region's data for one year. An attacker who could spoof telemetry or the price feed could choose the context the check judges against.

\section[Conclusion]{Conclusion\aimark}
\label{sec:conclusion}

Authorization by sender and familiar values misses attacks that use a legitimate credential and familiar values. Judging each write
against the feeder's operating context catches a large share of them at a
false-denial rate near its one-percent budget, and, in offline replay, catches
routine curtailment values issued at a positive price that a value check
allows. It complements static checks, and its coverage is limited to devices the operator can observe.

\section*{Acknowledgment}

Generative AI was used in every section. Anthropic Claude~\cite{aiClaude} wrote and revised the code and
drafted and edited the text; OpenAI GPT~\cite{aiGPT} and Google
Gemini~\cite{aiGemini} reviewed the code and manuscript; and a
literature-screening tool built on those models and on Qwen~\cite{aiQwen},
Granite~\cite{aiGranite}, Llama~\cite{aiLlama} and BGE~\cite{aiBGE} searched
and assessed related work. The author verified all AI-generated content and takes full responsibility for it. Every quantitative result was regenerated by running the code.

\bibliographystyle{IEEEtran}

\end{document}